\documentstyle[12pt,titlepage]{article}
\begin{document}
\begin{titlepage}
\addtolength{\baselineskip}{.7mm}
\begin{flushright}
NUP-A-96-15 \\
cond-mat/9611214\\
December, 1996
\end{flushright}
\begin{center}
{\large{\bf UNIVERSAL WIDE CORRELATORS
IN NON-GAUSSIAN ORTHOGONAL, UNITARY  AND SYMPLECTIC RANDOM MATRIX
ENSEMBLES}} \\[15mm]
\vspace*{3cm}
{\sc Chigak Itoi}
\footnote{\tt e-mail: itoi@phys.cst.nihon-u.ac.jp} \\[2mm]
{\it Department of Physics, \\[1mm]
College of Science and Technology, Nihon University, \\[1mm]
Kanda Surugadai, Chiyoda-ku, Tokyo 101, Japan} \\[6mm]
{\bf Abstract}\\[5mm]
{\parbox{13cm}{\hspace{5mm}
We calculate wide distance connected correlators in non-gaussian orthogonal,
unitary and symplectic random matrix ensembles by solving 
the loop equation in the $1/N$-expansion. 
The multi-level correlator  
is shown to be universal in large $N$ limit. 
We show the algorithm to obtain
the connected correlator to an arbitrary order
in the $1/N$-expansion.}}
\end{center}
\vspace*{5ex}
\end{titlepage}

\section{Introduction}
Recently,
the universality of  
large $N$ (wide) connected correlator in random matrix theories
is shown by Ambj{\o}rn, Jurkiewicz and Makeenko \cite{AJM}.
They solved the loop equation for the correlators 
with the motivation of 
studying the two dimensional gravity.
The function form of the wide 
correlator does not depend 
on the details of the probability distribution 
of a matrix ensemble. The universality classification in terms
of this correlator is also done in uniatry ensembles
by  Ambj{\o}rn and Akemann \cite{AA}. 
Br{\'e}zin and Zee stressed
this remarkably universal nature of random matrix theories
from the view point of the physics in disordered systems
\cite{BZ}. Although some other universal quantities in 
random matrix theories have been well-known already \cite{M},   
this new correlator can be calculated 
explicitly in various types of trace class ensembles
in several approaches, and its 
universal nature can be observed manifestly \cite{BHZ}.

In this paper, we calculate the large $N$  
correlators in non-gaussian orthogonal and symplectic 
ensembles 
by solving the loop equation. 
Ambj{\o}rn, Chekhov, Kristjansen and Makeenko 
developed this method originally proposed in \cite{AJM}
to calculate the correlator in principle
to an arbitrary order in $1/N$-expansion 
in an arbitrary distribution of a unitary invariant ensemble
\cite{ACKM}. Here, we extend this method to orthogonal 
and symplectic ensembles, which are well-known as  
useful description schemes for
the level statistics of time reversal non-integrable quantum 
systems without and with spin degrees of freedom, respectively
\cite{M}. 
Calculation scheme for wide connected correlator to an arbitrary order
in the $1/N$-expansion have never been obtained
in non-gaussian orthogonal and symplectic ensembles,
though its leading order is calculated by solving 
an integral equation \cite{B},
its universality is shown by the super matrix method \cite{HW}
and critical phenomena in an orthogonal ensemble
have been studied by Br{\'e}n and Neuberger 
as a model of two dimensional gravity on unoriented surface \cite{BN}. 
Here, we derive a loop equation of orthogonal
and symplectic ensembles in the eigenvalue representation
to calculate their wide connected correlator.
We find a different nontrivial term from the unitary case
in their loop equation.
At the leading order in the $1/N$-expansion,
the extension of the solving scheme 
is straightforward, while at the next to leading order
crucial modification is necessary.
Here we give a general scheme to solve the loop equation up to 
an arbitrary order of the $1/N$-expansion.

\section{The loop equation}
First we derive the loop equation for orthogonal($\beta=1$), 
unitary ($\beta=2$) and
symplectic ($\beta=4$) invariant ensembles. Their eigenvalue distribution
is given by
\begin{equation}
P(\lambda_1, \cdots, \lambda_N) 
= Z^{-1} |\Delta(\lambda_1, \cdots, \lambda_N )|^{\beta} 
\exp \left(- N \sum_{i=1} ^N V(\lambda_i) \right),
\end{equation}
where $Z$ is normalization constant
\begin{equation}
Z = \int_{-\infty} ^{\infty} d \lambda_1 \cdots \int_{-\infty} ^{\infty} 
d\lambda_N |\Delta(\lambda_1, \cdots, \lambda_N )|^{\beta} 
\exp \left(- N \sum_{i=1} ^N V(\lambda_i) \right), 
\label{Z}
\end{equation}
Van der Monde determinant
is defined by
$$
\Delta(\lambda_1, \cdots, \lambda_N) = 
\prod_{1 \leq i< j \leq N}  (\lambda_i - \lambda_j),
$$
and the potential is a polynomial 
$V(x) = \sum_{k = 1} ^n \frac{g_k}{k} x^k $.
The Jacobian of an infinitesimal transformation of the integration variables 
$\lambda_i ^{\prime} =  \lambda_i + \frac{\epsilon}{z- \lambda_i}$
gives
\begin{eqnarray}
&d^N \lambda^{\prime}& |\Delta(\lambda_1 ^{\prime}, \cdots, \lambda_N ^{\prime} )|^{\beta} \nonumber \\
 = 
&d^N\lambda& |\Delta (\lambda_1, \cdots, \lambda_N )|^{\beta}
\left( 1+ \epsilon \left( \frac{\beta}{2} 
( \sum_{i=1} ^N \frac{1}{z - \lambda_i} )^2
 + (1- \frac{\beta}{2}) \sum_{i=1} ^N  \frac{1}{(z-\lambda_i)^2} \right) \right).
\end{eqnarray}
This transformation in the integral (\ref{Z}) gives the loop equation
\begin{equation}
\frac{\beta}{2} \langle \left(\sum_{i=1} ^N \frac{1}{z-\lambda_i} 
\right)^2 \rangle  +
\left( \frac{\beta}{2} - 1 \right)\partial_z \langle \sum_{i=1} ^N \frac{1}{z-\lambda_i}\rangle
-N \langle \sum_{i=1} ^N V^{\prime}(\lambda_i)\frac{1}{z-\lambda_i} \rangle 
=0,  
\label{loop1}
\end{equation}
where $\langle \cdots \rangle = 
\int d^N \lambda (\cdots) P(\lambda_1, \cdots, \lambda_N)$.
Here, we define one point green function
$$
G(z)= \langle \frac{1}{N} 
\sum_{i=1} ^N \frac{1}{z-\lambda_i} \rangle
$$
and $l$-level connected correlator
$$
G(z_1, \cdots, z_l)=N^{l-2} \langle
\sum_{i_1=1} ^N \frac{1}{z_1 -\lambda_{i_1}} \cdots
\sum_{i_l=1} ^N \frac{1}{z_l-\lambda_{i_l}} \rangle_c.
$$ 
As discussed by Ambj{\o}rn, Chekhov Kristjansen and Makeenko \cite{ACKM}, the
loop equation (\ref{loop1}) can be rewritten in terms of the one point green function and the connected two level corelator as the following closed form
\begin{equation}
G(z)^2 -\frac{2}{\beta} \oint_C \frac{d w}{2 \pi i} 
\frac{V^{\prime}(w)}{z-w} G(w) 
-\frac{1}{N} \left(\frac{2}{\beta} - 1 \right) \partial_z G(z)
+ \frac{1}{N^2} G(z, z) = 0,
\label{loop2}
\end{equation}
where the contour $C$ encircles the singularities of $G(z)$. 
In the case of a finite $N$, these must consist of pole singularities,
while in the $1/N$-expansion these make up cuts. Here we assume
the coefficient in an arbitrary order has cuts 
$(x_1, x_2), \cdots, (x_{2s-1}, x_{2s})$.
The third term in the left hand side in eq(\ref{loop2}) 
is absence in the unitary case. This term requires a
necessary modification of the 
$1/N$-expansion 
$$
G(z_1, \cdots, z_l ) = \sum_{n=1} ^{\infty} 
\frac{1}{N^n} G_n(z_1, \cdots, z_l). 
$$ 
Note that there is no odd degree term in the unitary ensemble.
The coefficient $G_n(z)$ and $G_n(z, z)$ of an arbitrary order
satisfies the following recursion relation
\begin{eqnarray}
&n=0,&
\hat{X} G_0 (z)=0, 
\label{loopG0} \\
&n=1,&
\hat{X}G_1(z)= \left(1- \frac{2}{\beta} \right) \partial_z G_0(z),
\label{loopG1} \\
&n \geq 2,& \hat{X} G_n(z)= \sum_{m=1} ^{n-1} G_{n-m}(z) G_m(z)
+ \left(1- \frac{2}{\beta} \right) \partial_z G_{n-1}(z)
+ G_{n-2}(z, z),
\label{loopGn}
\end{eqnarray}
where we define the operator $\hat{X}$ acting 
on an arbitrary function $f(z)$ by
\begin{equation}
\hat{X}f(z)\equiv
\frac{2}{\beta}\oint_C \frac{d w}{2 \pi i} 
\frac{V^{\prime}(w)}{z-w} f(w) - 2 G_0(z) f(z).
\label{X}
\end{equation}
We shall derive large $N$ correlators by solving this equation
recursively. 

\section{Large $N$ limit}
Here we obtain $G_0(z)$, $G_0(z, w)$ and $G_0(p, q, r)$
in a non-gaussian ensemble.
In the leading order $n=0$, the equation for $G_0$ can be written as
\begin{equation}
\frac{\beta}{2} G_0(z) ^2
- V^\prime (z) G_0(z) -R(z, V) = 0,
\label{loop0}
\end{equation}
where $R(z, V)$ is defined by  
$$
\frac{1}{N} \langle \sum_{i=1} ^N 
V^{\prime}(\lambda_i)\frac{1}{z-\lambda_i} \rangle 
= V^\prime (z) G(z) + R(z, V). 
$$
The quadratic equation (\ref{loop0}) is solved with respect to $G_0$
\begin{equation}
G_0(z) = \frac{1}{\beta} \left( V^\prime (z) 
-\sqrt{V^\prime (z) ^2  +2 \beta R(z, V)}\right).
\label{sol0}
\end{equation}
If we assume s-cut solution, $G_0$ should have the form with the 
certain polynomial $M(z)$. 
\begin{equation}
G_0(z) = \frac{1}{\beta}\left( V^\prime (z) -M(z) 
\sqrt{ \prod_{i=1} ^{2s}(z-x_i) }\right).
\label{sol1}
\end{equation}
This assumption gives the following integral expression for $M$
$$
M(z) = \oint_{C_\infty} \frac{dw}{2 \pi i}\frac{1}{w-z} M(w)
= \oint_{C_\infty} \frac{dw}{2 \pi i} 
\frac{V^\prime(w)- \beta G_0(w)}{(w-z) \sqrt{\prod_{i=1} ^{2s} (w-x_i)}},
$$
where the contour $C_\infty$ is a circle with the infinitely large radius. 
The second term of the integral vanishes due to the 
asymptotic behavior of $G_0(z) \sim 1/z$
for $|z| \sim \infty$ and therefore 
$$
M(z)= \oint_{C_\infty} \frac{dw}{2 \pi i} 
\frac{V^\prime (w)}{(w-z) \sqrt{\prod_{i=1} ^{2s}(w-x_i)}}.
$$  
Substituting this into eq(\ref{sol1}), one obtains 
\begin{eqnarray}
G_0(z) &=& 
\frac{1}{\beta} \left( V^\prime (z) -
\oint_{C_\infty} \frac{dw}{2 \pi i}\frac{V^\prime (w)}
{w-z} \sqrt{\prod_{i=1} ^{2s}
 \frac{z-x_i}{w-x_i}} \right) \\
&=& -\frac{1}{\beta} \oint_C \frac{dw}{2 \pi i}\frac{V^\prime (w)}{w-z} 
\sqrt{\prod_{i=1} ^{2s}
 \frac{z-x_i}{w-x_i}} \label{1pg} 
\end{eqnarray} 
The large $z$ asymptotics of $G_0$ in this expression gives
\begin{equation}
\oint_C \frac{dw}{2 \pi i} \frac{w^r V^\prime(w)}{\sqrt{\prod_{i=1} ^{2s}
(w-x_i)}} = \delta_{r s}, \ \ \ \ r=0, \cdots, s.
\label{asympt}
\end{equation}
These equations and the arc independence of the chemical potencial
\begin{equation}
0 = \int_{x_{2k}} ^{x_{2k+1}} d \lambda M(\lambda) 
\sqrt{\prod_{i=1} ^{2s}(\lambda-x_i)},
\ \ \ k=1, \cdots, s-1,
\label{chemical}
\end{equation}
determine the location of the edges 
$x_{k},\ (k=1, \cdots, 2s)$ of the cuts \cite{J}.
In the $s=1$ case,
the edge values $x_1, \ x_2$ are determined only by the 
condition (\ref{asympt}).
Here we calculate a wide two level correlator in the $s=1$ case
in the large $N$ leading order.
The green function is obtained by acting the loop
insertion operator
\begin{equation}
\frac{d}{dV(w)} \equiv -\sum_{j=1} ^\infty \frac{j}{w^{j+1}}
\frac{d}{dg_j},
\label{insertion}
\end{equation}
on the free energy $F \equiv \frac{1}{N^2}\log Z$
\begin{equation}
G(z) = \frac{dF}{dV(z)} + \frac{1}{z}.
\label{1green}
\end{equation}
The $l (\geq2)$ level correlator is obtained as well
\begin{equation}
G(z_1, \cdots, z_l) =  \frac{d}{dV(z_1)} \cdots \frac{d}{dV(z_l)} F.
\label{2level}
\end{equation}
Thanks to the formula
$$
\frac{dV^\prime (z)}{dV(w)} = \frac{-1}{(z-w)^2}, 
$$
we can differentiate the one point 
green function $G_0(z)$ given by eq(\ref{1pg}) to calculate 
the two level correlator in the leading order of the $1/N$-expansion
\begin{eqnarray}
G_0(p, q) 
&=& - \frac{d}{dV(q)}\frac{1}{\beta} \oint_C \frac{dw}{2 \pi i}\frac{V^\prime (w)}{w-p} 
\sqrt{\prod_{i=1} ^{2}
 \frac{p-x_i}{w-x_i}} 
\nonumber \\
&=& \frac{1}{\beta} \oint_C \frac{dw}{2 \pi i} 
\frac{-1}{(w-q)^2} \frac{1}{p-w} 
\sqrt{\frac{(p-x_1)(p-x_2)}{(w-x_1)(w-x_2)}} \nonumber \\
&+& \frac{1}{\beta} \oint_C \frac{dw}{2 \pi i} \frac{V^\prime(w)}{p-w}
\sqrt{\frac{(p-x_1)(p-x_2)}{(w-x_1)(w-x_2)}}\frac{1}{2} \sum_{k=1} ^{2s}
\left(\frac{1}{w-x_k} - 
\frac{1}{p-x_k} \right) \frac{dx_k}{dV(q)} \nonumber \\
&=& \frac{1}{\beta} \left[ \frac{-1}{(p-q)^2} + 
\frac{\partial}{\partial q}\left( \frac{1}{p-q} 
\sqrt{\frac{(p-x_1)(p-x_2)}{(q-x_1)(q-x_2)}} \right) \right] \nonumber \\
&+& 
\frac{1}{\beta} \left[ \frac{1}{2} M_1 ^{(1)} \frac{d x_1}{dV(q)}
\sqrt{\frac{p-x_2}{p-x_1}} 
+ \frac{1}{2} M_2 ^{(1)} \frac{d x_2}{dV(q)}
\sqrt{\frac{p-x_1}{p-x_2}} \right],
\label{2level2}
\end{eqnarray}
where the moment $M_i ^{(k)}$ is defined by
$$
M_i ^{(k)} \equiv \oint_C \frac{dw}{2 \pi i} \frac{V^\prime(w)}{(w-x_i)^k
\sqrt{(w-x_1)(w-x_2)}}.
$$
In this single arc case, $M_1 ^{(1)} \frac{dx_1}{dV(q)}$ and $M_2 ^{(1)} \frac{dx_2}{dV(q)}$
is obtained only by differentiating the condition (\ref{asympt})
\begin{eqnarray}
M_1 ^{(1)} \frac{dx_1}{dV(q)} &=& \frac{1}{(q-x_1)\sqrt{(q-x_1)(q-x_2)}} 
\nonumber \\  M_2 ^{(1)} \frac{dx_2}{dV(q)} &=& \frac{1}{(q-x_2)\sqrt{(q-x_1)(q-x_2)}}.
\label{edge}
\end{eqnarray}
Substituting these into eq(\ref{2level2}), 
we obtain the following expression of the wide two level correlator
\begin{equation}
G_0(p, q) = \frac{pq - (x_1+x_2)(p+q)/2 + x_1 x_2}{\beta 
(p-q)^2 \sqrt{(p-x_1)(p-x_2)(q-x_1)(q-x_2)}} -\frac{1}{\beta(p-q)^2}.
\label{2levelng}
\end{equation}
This result agrees with that in \cite{AJM} in the 
unitary case $\beta=2$  and that in \cite{B}
for an arbitrary $beta$.
We can calculate the three level correlator from this expression 
and the derivative of the edge values (\ref{edge}) 
\begin{equation}
G_0(p, q, r) = \frac{(x_2-x_1)[(p-x_2)(q-x_2)(r-x_2)/M_1 ^{(1)}-
 (p-x_1)(q-x_1)(r-x_1)/M_2 ^{(1)}]}
{4 \beta [(p-x_1)(q-x_1)(r-x_1)(p-x_2)(q-x_2)(r-x_2)]^{3/2}}
\label{3levelng}
\end{equation}
This result
agrees with those obtained in \cite{AJM} in the 
unitary case.
 It is obvious that the
correlator of the level more than two takes a universal form,
since the two level correlator depends on the distribution
$V(\lambda)$ only through the edge values

\section{Gaussian ensemble}
  Here we concern ourselves with the gaussian distribution.  
We calculate the one point green function up to second order in the $1/N$-expansion.
In the gaussian model, the one point green function and 
the wide two level correlator become
\begin{eqnarray}
&& G_0(z) = \frac{1}{\beta} \left(z- \sqrt{z^2 - 2 \beta} \right) 
\label{gauss1} \\
&&G_0(z, w) = \frac{zw-2 \beta}
{\beta(z-w)^2 \sqrt{(z^2- 2 \beta)(w^2 - 2 \beta)}}
-\frac{1}{\beta(z-w)^2}. \label{gauss2}
\end{eqnarray} 
The loop equation (\ref{loopG1}) for the next order $G_1(z)$ is
calculated by an reasonable assumption that 
the green function has no singularity outside 
the contour $C$ and the asymptotic behavior 
$G_1(z) \sim 1/z^2$ for $|z| \sim \infty$
give
$$
\oint_C \frac{dw}{2 \pi i} \frac{w}{z-w} G_1(w) = z G_1 (z),
$$
which yields the special solution of the loop equation (\ref{loopG1})
\begin{equation}
G_1(z)= \frac {1}{2}\left( 1- \frac{2}{\beta} \right) \left( \frac{1}{\sqrt{z^2-2 \beta}} 
-\frac{z}{z^2 - 2 \beta} \right).
\label{special1}
\end{equation}
The general solution is given by adding the general solution of the homogeneous 
equation with respect to $G_1(z)$ 
$$
\frac{c_1}{\sqrt{z^2-2 \beta}} + \frac{c_2 z}{\sqrt{z^2-2 \beta}},
$$
which does not satisfy the large $z$ asymptotics. Therefore the unique solution 
of eq(\ref{loopG1}) with the correct asymptotic behavior is  given by eq(\ref{special1}). These results eq(\ref{gauss1}),
eq(\ref{gauss2}) and eq(\ref{special1}) agree with those obtained
by the replica trick \cite{IMS}. \\

The loop equation (\ref{loopGn}) for $G_2(z)$ is 
\begin{equation}
\hat{X}G_2( z)= \left(1-\frac{2}{\beta}\right) \partial_z G_1(z)
+G_1(z)^2 +G_0(z, z),
\label{loopG2}
\end{equation}
The two level correlator with the coincident point can be defined 
as a finite function by limiting procedure 
$$
G_0(z, z) \equiv \lim_{w \rightarrow z} G_0(z, w) = \frac{1}{(z^2-2 \beta)^2},
$$
which enables us to derive the solution of eq(\ref{loopG2})
\begin{equation}
G_2(z) = \frac{(2-\beta)^2 (2 z^2+\beta -2 z \sqrt{z^2-2 \beta}) +2 \beta^2}
{4 \beta(z^2-2\beta)^{5/2}}.
\end{equation}
This is the unique solution of eq(\ref{loopG2}) as derived $G_1(z)$. \\

\section{Higher orders in the $1/N$-expansion}
Now, we describe a general algorithm to 
solve eq(\ref{loopGn}) to an arbitrary order
in non-gaussian case. In the orthogonal $\beta=1$ and 
symplectic $\beta=4$ cases, the derivative term 
$\partial_z G_{n-1}(z)$ in the right hand side of 
eq(\ref{loopG1}) and eq(\ref{loopGn}) requires 
the non-trivial modification of the algorithm 
in the unitary $\beta=2$ case.
We have to solve the following type of 
equation for an unknown function
$G(z)$ and a given function $F(z)$
\begin{equation}
\hat{X}G(z)= F(z),
\label{inteq}
\end{equation}
We assume 
the regularity of $G(z)$, $G_0(z)$ and $F(z)$ outside the region encircled
by the contour $C$ and the large $z$ asymptotics of $G_0(z) \sim 1/z$
and  $G(z) \sim 1/z^2$. 
A rewriting of the second term in the operation of $\hat{X}$
and a deformation of the contour $C_z$ to $C$ and $C_\infty$
give
\begin{eqnarray}
G_0(z) G(z) &=& \oint_{C_z} \frac{dw}{2 \pi i} \frac{G_0(w)G(w)}{w-z}
=-\left( \oint_{C} + \oint_{C_\infty} \right) 
\frac{dw}{2 \pi i} \frac{G_0(w)G(w)}{w-z} \nonumber \\
&=& \oint_{C} \frac{dw}{2 \pi i} \frac{G_0(w)G(w)}{z-w}.
\end{eqnarray}
The contribution from the contour integral of $C_\infty$ vanishes
due to the large $z$ asymptotics of $G_0(z)$ and $G(z)$.
Therefore the operation of $\hat{X}$ can be written as
\begin{equation}
\hat{X}G(z)
=2 \oint_C \frac{d w}{2 \pi i} 
\frac{M(w)\sqrt{\prod_{i=1} ^{2s} (w-x_i)}}{z-w} G(w).
\label{OP}
\end{equation}
In this expression of the operator $\hat{X}$, we can show a special solution of eq(\ref{inteq}). If $M(z)^{-1}$ has no pole outside the region encircled the contour $C$, we find a special solution 
\begin{equation}
G^S(z) = \frac{F(z)}{M(z) \sqrt{\prod_{i=1} ^{2s}(z-x_i)}}.
\label{spsol1}
\end{equation}
In general, $M(z)^{-1}$ has poles 
outside the contour $C$. We have to eliminate these 
unphysical singularities from the expression. Therefore,
the solution becomes
\begin{eqnarray}
G^S(z) &=& \frac{F(z)}{2 M(z) \sqrt{\prod_{i=1} ^{2s}(z-x_i)}}
-\oint_{C \cup C_z} \frac{dw}{4 \pi i} \frac{F(w)}
{(w-z)M(w)\sqrt{\prod_{i=1} ^{2s}(z-x_i)}},
\label{spsol2} \\ &=&
\oint_{C} \frac{dw}{4 \pi i} \frac{F(w)}
{(z-w)M(w)\sqrt{\prod_{i=1} ^{2s}(z-x_i)}}
\end {eqnarray}
which has no pole outside the contour $C$.
Since the second term in the right 
hand side in eq(\ref{spsol2})is annihilated by
the action of the operator $\hat{X}$ with the expression eq(\ref{OP}),
this $G^S(z)$ satisfies eq(\ref{inteq}).
General solution $G(z)$ of eq(\ref{inteq}) is 
given by the linear combination of this special solution $G^S (z)$ and 
the allowed zero modes of the operator $\hat{X}$. 
In $s \geq 2$ case, Akemann shows that the allowed zero modes are
$$
\frac{z^0}{\sqrt{\prod_{i=1}^{2s}(z-x_i)}}, \cdots,
\frac{z^{s-2}}{\sqrt{\prod_{i=1}^{2s}(z-x_i)}}, 
$$
which obey the boundary condition $G(z) \sim 1/z^2 (z \rightarrow \infty )$ 
\cite{A}. In $s=1$ case, the boundary condition excluded any zero mode
to be added to $G^S (z)$. The unique solution of $G(z)$ as 
the green function is determined by the condition eq(\ref{1green}) that 
the green function which is
the total derivative with respect to the loop insertion 
operator $\frac{d}{dV(z)}$.
The equation (\ref{loopGn}) for $G_n(z)$ 
can be solved recursively
in the above shown formula, and a correlator with an arbitrary level
can be calculated to an arbitrary order in $1/N$-expansion 
by acting the loop insersion operator eq(\ref{insertion}) on the one point green function $G_n(z)$. 

\section{Summary}
We have calculated two level and three level correlators
in the large $N$ limit in non-gaussian orthogonal, unitary and 
symplectic ensembles. We have obtained  
explicit forms of higher orders $G_1(z)$ and $G_2(z)$ in
the gaussian ensemble. These results
are consistent with those obtained in other method \cite{IMS}.
We have shown a new general algorithm to solve the
loop equation recursively to
calculate a correlator up to an arbitrary order in the $1/N$-expansion
for non-gaussian orthogonal and symplectic ensembles. \\

I would like to thank S. Higuchi, H. Mukaida and Y. Sakamoto
for helpful discussions.  
I am grateful to S. Hikami for explaining his recent work.
\\

\end{document}